\documentclass[aps,prl,twocolumn,groupedaddress]{revtex4}
\usepackage{graphicx}

\begin{document}

\title{Onset of Interlayer Phase Coherence in a Bilayer Two-Dimensional Electron System: Effect of Layer Density Imbalance}

\author{I.~B. Spielman$^1$, M. Kellogg$^1$, J.~P. Eisenstein$^1$, L.~N. Pfeiffer$^2$, and K.~W. West$^2$}

\affiliation{$^1$California Institute of Technology, Pasadena CA 91125 
\\
	 $^2$Bell Laboratories, Lucent Technologies, Murray Hill, NJ 
07974\\}

\date{\today}

\begin{abstract}
Tunneling and Coulomb drag are sensitive probes of spontaneous interlayer phase coherence in bilayer two-dimensional electron systems at total Landau level filling factor $\nu_T = 1$.  We find that the phase boundary between the interlayer phase coherent state and the weakly-coupled compressible phase moves to larger layer separations as the electron density distribution in the bilayer is imbalanced. The critical layer separation increases quadratically with layer density difference. 

\end{abstract}


\maketitle
Bilayer two-dimensional electron systems (2DES) are most interesting when the separation between the layers is comparable to the average distance between electrons in the individual layers. Interlayer Coulomb interactions are then just as important as interlayer ones and the system supports collective phases that do not exist in the individual layers\cite{perspectives}.  A particularly interesting example occurs when a magnetic field $B$ is applied perpendicular to the layers and the total density $N_T = N_1+N_2$ of electrons in the system equals the degeneracy $eB/h$ of the lowest spin-resolved Landau level produced by the field. In this case the total Landau level filling factor $\nu_T = \nu_1+\nu_2 =hN_T/eB = 1$. Beyond exhibiting a quantized Hall effect (QHE) when electrical currents flow in parallel through the two layers, this many-body phase displays a variety of fascinating phenomena associated with observables which are antisymmetric in the layer degree of freedom.  For example, a giant enhancement of the zero bias interlayer tunneling conductance has been observed\cite{spielman}, as has the vanishing of both the longitudinal {\em and} Hall resistances of the system when equal but oppositely directed currents flow in the two layers\cite{kellogg,tutuc}.  These findings strongly support the idea that the ground state of the system is a Bose condensate of phase coherent interlayer excitons. 

As the separation between the layers is increased, the excitonic phase weakens and a poorly-understood transition to a state exhibiting none of the above properties occurs.  At very large layer separation a {\it balanced} bilayer system (i.e. one in which $N_1 = N_2$ and thus $\nu_1 = \nu_2 = 1/2$) may be regarded as two independent compressible composite fermion liquids\cite{halperin}. However, near the critical layer separation the situation is much less clear. Indeed, it is not known what the order of the transition is nor whether intermediate phases exist between the excitonic phase and the weakly-coupled composite fermion fluid. Recent experiments\cite{kellogg2} have shown a strong enhancement of the longitudinal Coulomb drag in the transition region which has been interpreted\cite{ady} in terms of phase separation induced by static disorder.

If the bilayer system is imbalanced, but remains at $\nu_T =1$, the spectrum of possible phases widens further. At large layer separation, both compressible (e.g. $\nu_1 = 1/4$, $\nu_2 = 3/4$) and incompressible (e.g. $\nu_1 = 1/3$, $\nu_2 = 2/3$) possibilities exist.  At small separations, deep in the excitonic phase, small layer imbalances are not expected to be qualitatively important, since this state is characterized by a broken U(1) symmetry in which fluctuations in the layer density difference $N_1-N_2$ are large. Larger imbalances may, however, lead to the defeat of the excitonic phase by competing phases, with possibilities including one or more conjugate pairs of fractional quantized Hall states such as $(\nu_1,\nu_2)$ = (1/3, 2/3); (2/5, 3/5); (3/7, 4/7); etc.  In this paper we report interlayer tunneling and Coulomb drag measurements which clearly indicate that near the critical layer separation small layer density imbalances enhance the stability of the interlayer phase coherent excitonic phase. A quantitative determination of the shape of the phase boundary for small imbalances is presented.

The samples used in these experiments are GaAs/AlGaAs double quantum wells grown by molecular beam epitaxy. Two 18 nm GaAs quantum wells are separated by a 10 nm $\rm Al_{0.9}Ga_{0.1}As$ barrier. Remote Si dopants yield 2DES's in each well with densities of about $5.5 \times 10^{10} \rm cm^{-2}$ and mobilities around $1.0 \times 10^{6} \rm cm{^2}/Vs$. The splitting between the lowest symmetric and antisymmetric states in the double well potential is estimated to be $\Delta_{SAS} \approx 90 \mu$K. Separate electrical contacts to the individual 2D layers are realized using a local selective depletion technique. For tunneling studies square mesas 250 $\mu$m on a side (with four contact arms extending outward) are patterned onto the wafer, while for Coulomb drag experiments a rectangular mesa (160 by 320 $\mu$m) with seven contact arms is used. Metal gates deposited on the front and back sides of the central mesa region provide for independent control of the layer densities $N_1$ and $N_2$ and thus allow for the creation of both balanced and imbalanced bilayer systems.  The action of these gates is calibrated via measurements of the low magnetic field quantum oscillations of the interlayer tunneling and/or resistivity of the individual layers. Control over the total electron density also allows for continuous tuning of the ratio of intra- to interlayer Coulomb interactions in the sample. This ratio is conveniently parameterized by $d/\ell$, with $d$ = 28 nm being the center-to-center quantum well separation and $\ell = (\hbar/eB)^{1/2}$ the magnetic length. Prior experiments\cite{spielman,kellogg} have shown that the transition from the weakly-coupled compressible phase to the strongly-coupled excitonic phase at $\nu_T = 1$ occurs near $d/\ell \sim 1.7 - 1.8$ in the present samples. Since the ratio of the tunnel splitting $\Delta_{SAS}$ to the mean Coulomb energy $e^2/\epsilon\ell$ is only about $1 \times 10^{-6}$, the $\nu_T = 1$ QHE in these sample is overwhelmingly dominated by electron-electron interactions. 

Figure 1a shows two representative low temperature ($T$ = 25 mK) interlayer tunneling conductance spectra at $\nu_T =1$ in a balanced bilayer 2DES. The lower trace was obtained at $d/\ell = 1.82$ and is representative of the compressible phase just above the critical layer separation.  Near zero interlayer voltage $V$ the conductance $dI/dV$ is heavily suppressed\cite{jpe}.  This is a single layer effect and reflects the fact that at high magnetic field a 2DES is strongly correlated, irrespective of whether it is compressible or incompressible.  The sudden injection, via tunneling, of an {\it uncorrelated} electron into the 2DES can only produce a highly excited state; no low energy states are accessible on the rapid time-scale of the tunneling event. The upper trace in Fig. 1a was obtained at $d/\ell = 1.76$, inside the strongly-coupled excitonic phase.  Instead of a suppression of tunneling at zero bias, there is now a huge and very sharply resonant enhancement.  This Josephson-like effect has been widely interpreted as a direct result of spontaneous interlayer phase coherence in the bilayer system.  The strong interlayer correlations ensure that an electron about to tunnel always faces a hole in the opposite layer and is thus fully correlated in advance. 

\begin{figure}
\centering
\includegraphics[width=3.2 in, bb=112 407 374 748]{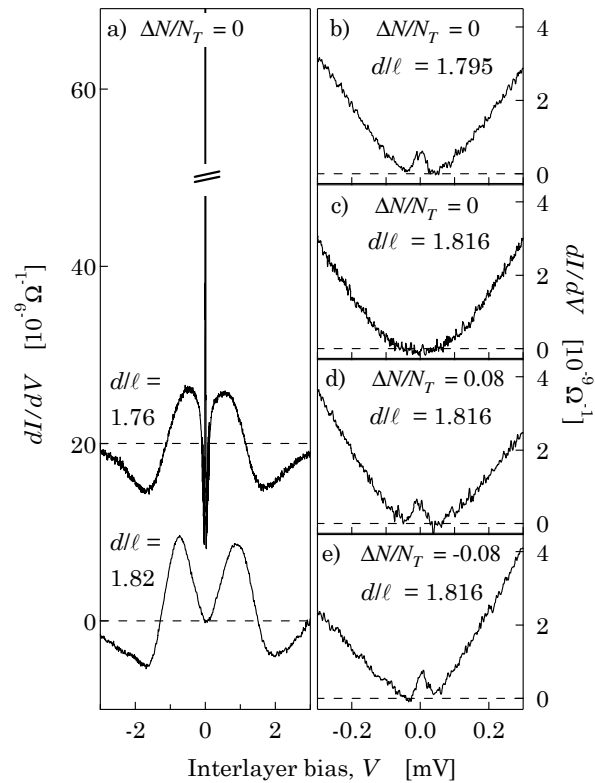}
\caption{\label{fig:fig1}Tunneling spectra at $\nu_T = 1$ and $T = 25$mK. a) Balanced bilayer. Upper trace: $d/\ell = 1.76$, coherent phase. Lower trace: $d/\ell = 1.82$, compressible phase. Upper trace shifted vertically for clarity. b) Enlargement of zero bias region at $d/\ell = 1.795$, just inside the coherent phase in the balanced case. A small peak near zero bias is evident. c), d) and e) Spectra at $d/\ell = 1.816$ in both balanced and imbalanced cases.  While no zero bias peak is seen at balance, the small density imbalance creates one. }
\end{figure}

The remaining panels of Fig. 1 show enlargements of the zero bias region of tunneling spectra from the same sample but with $d/\ell$ very close to the critical value separating the weakly- and strongly-coupled phases. Figs. 1b and 1c contain $\nu_T = 1$ tunneling spectra in the balanced configuration at $d/\ell = 1.795$ and 1.816 respectively.  The data in Fig. 1b show a weak peak near zero bias, demonstrating that the sample is just inside the excitonic phase.  In Fig. 1c the peak is absent; the sample is evidently just outside the excitonic phase\cite{noise}.  Most interesting are Figs. 1d and 1e.  Here the total density $N_T$ is the same as in Fig. 1c and thus $d/\ell = 1.816$, but now the sample is imbalanced: $N_1 = (N_T + \Delta N)/2$ and $N_2 = (N_T - \Delta N)/2$. In Figs. 1d and 1e $\Delta N/N_T = \pm 0.08$, respectively\cite{shift}. In both cases a peak has appeared at zero bias. The data in Figs. 1c, 1d, and 1e convincingly demonstrate that a small layer density imbalance can stabilize the excitonic phase even when it is not present in the balanced configuration at the same total density. 

A similar imbalance-induced stabilization of the strongly-coupled $\nu_T = 1$ excitonic phase is observed in Coulomb drag experiments.  In such measurements a current $I$ driven through one of the 2D layers produces voltage drops $V_D$ in the other layer\cite{tunneldrag}. At zero magnetic field the drag voltage is parallel to the current and is simply proportional to the interlayer momentum relaxation rate.  Deep within the $\nu_T = 1$ excitonic phase the quantum Hall energy gap suppresses inelastic interlayer Coulomb scattering events and the observed longitudinal drag voltage at low temperatures is exponentially small\cite{kellogg3}.  However, a strong transverse, or Hall, component of the drag is observed. In the strongly-coupled phase the Hall drag resistance $R_{xy,D} = V_{xy,D}/I$ is in fact precisely quantized, with $R_{xy,D} = h/e^2$\cite{kellogg3}. Strong Hall drag is unusual and is believed to be a direct signature of non-trivial interlayer correlations\cite{renn,duan,yang1,yang2,kim}. Girvin\cite{girvin} has offered an intuitive explanation of the effect. In the strongly-coupled bilayer $\nu_T = 1$ phase, each electron ``sees" a vortex, or node, in the many-body wavefunction at the location of each of the other electrons, {\it irrespective of which layer they are in}.  Thus, a current flowing solely in one layer produces a flow of vortices in the other layer.  Via the Josephson relation, this flow of vortices produces a transverse voltage in the non-current-carrying layer.  This is Hall drag and $R_{xy,D} = h/e^2$ follows immediately.
 
\begin{figure}
\centering
\includegraphics[width=3.0 in,bb=19 631 238 771]{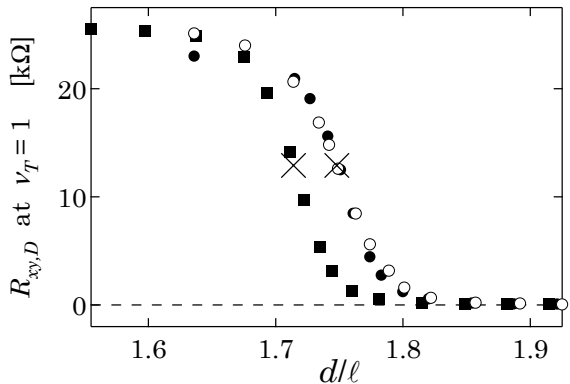}
\caption{\label{fig:fig2} Hall drag vs. $d/\ell$ at $\nu_T = 1$ and $T = 50$ mK. Solid squares: Balanced bilayer, $\Delta N/N_T = 0$. Solid and open circles: Imbalanced bilayer, $\Delta N/N_T = \pm 0.1$.}
\end{figure}

Recent experiments\cite{kellogg2} have shown that $R_{xy,D}$ at $\nu_T = 1$ rises from zero to the quantized value over a fairly narrow range of effective layer separations $d/\ell$ about the critical value separating the two phases.  Figure 2 shows Hall drag data at $\nu_T = 1$ and $T = 50$ mK from a bilayer sample taken from the same parent wafer as the tunneling sample described above. These data were obtained by first measuring the magnetic field dependence of $R_{xy,D}$ at fixed layer densities and then picking out the value at $\nu_T = 1$.  The transition of $R_{xy,D}$ at $\nu_T = 1$ from 0 to $h/e^2$ as $d/\ell$ is reduced is shown both in balanced $N_1 = N_2$ and imbalanced ($\Delta N/N_T = \pm 0.10$) situations.  It is clear from the figure that imbalance causes the midpoint of the drag transition to move to larger effective layer separations. 

Figure 3 shows the boundaries in the $d/\ell$ vs. $\Delta N/N_T$ plane which separate the weakly-coupled and strongly-coupled phases at $\nu_T = 1$ as deduced from tunneling and drag data like those in Figs. 1 and 2.  For tunneling, sets of conductance spectra at fixed total density $N_T$ (and thus fixed $d/\ell$) but various layer imbalances $\Delta N/N_T$ were examined for the presence of a zero bias peak. For those densities and effective layer separations where the peak was absent at balance but present at sufficiently large imbalance, a point on the phase boundary could be determined.
For Coulomb drag, the critical $d/\ell$ value at a given imbalance $\Delta N/N_T$ was taken to be that where $R_{xy,D} = 0.5h/e^2$.  It is clear from the figure that both tunneling and Hall drag suggest that the critical layer separation $d/\ell$ rises roughly quadratically with $\Delta N/N_T$.

\begin{figure}
\centering
\includegraphics[width=3.0 in,bb=16 600 237 771]{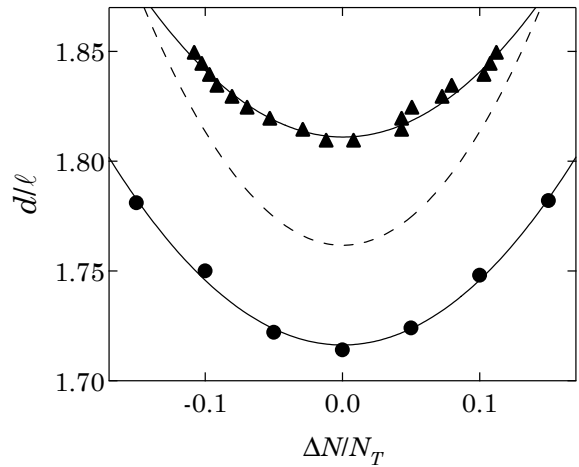}
\caption{\label{fig:fig3} Boundaries in $d/\ell$ vs. $\Delta N/N_T$ separating the weakly-coupled and strongly-coupled phases at $\nu_T = 1$ as deduced from tunneling (triangles) and Hall drag (dots)\cite{shift}. Solid lines are least squares fits to parabolas.  Dashed curve is modified theoretical prediction of Ref.\cite{yogesh}.}
\end{figure}

The phase boundaries determined by tunneling and Hall drag shown in Fig. 3 are displaced from one another by approximately $\Delta(d/\ell) = 0.1$, in spite of
the fact that the data were obtained with samples from the same semiconductor wafer.  This difference is in large part a result of the way we have defined the transition point in the two experiments. In Hall drag the phase boundary is identified with the $d/\ell$ value where $R_{xy,D}$ reaches one-half of its quantized value $h/e^2$.  As Fig. 2 makes clear, non-zero Hall drag is observed at noticeably larger layer separations than this.  In contrast, the tunneling transition is identified by the first observation of a tiny zero bias peak.  This is the only sensible definition in this case, since unlike Hall drag, the tunneling conductance is not changing from zero to some universal value as the phase boundary is crossed.  Beyond this, we note that Hall drag and tunneling may depend very differently on the connectivity of regions in the sample in which the strongly-coupled phase exists.  Stern and Halperin\cite{ady} have argued that static fluctuations in the layer densities of real samples lead to phase separation in the transition region.  As $d/\ell$ is reduced toward the phase boundary, initially only small regions of the strongly-coupled excitonic phase appear within the weakly-coupled background fluid. While interlayer tunneling will detect these regions almost immediately, the quantization of the drag and conventional Hall resistances requires them to percolate which will only occur at smaller $d/\ell$. 

Several prior experiments have suggested that the QHE in bilayer $\nu_T = 1$ systems is robust against layer density imbalance\cite{sawada,dolgopolov,tutuc2,clarke}.  Sawada, {\it et al.}\cite{sawada} reported that the strength of the $\nu_T = 1$ QHE (inferred from the width of the quantized Hall plateau  and the energy gap $\Delta$ extracted from the temperature-dependent diagonal resistivity) increased symmetrically with layer density imbalance.  These measurements, however, were made in a sample with very strong tunneling (the symmetric-antisymmetric tunnel splitting was $\Delta_{SAS} \approx 7$K, considerably larger than the observed transport gap $\Delta$) and thus were not necessarily representative of the physics of $\nu_T = 1$ QHE in the Coulomb-dominated, spontaneously interlayer phase coherent regime. Tutuc, {\it et al.}\cite{tutuc2}, using strongly correlated 2D hole bilayers with very weak interlayer tunneling, also found the energy gap of the $\nu_T = 1$ QHE to increase with layer imbalance.  Finally, Clarke, {\it et al.}\cite{clarke}, again using bilayer hole samples, found that the width of the $\nu_T = 1$ QHE resistivity minimum either remained the same as the sample was imbalanced, or increased.  Clarke, {\it et al.} also claimed that the interlayer phase coherent $\nu_T = 1$ state could develop in the presence of layer density imbalance even when it is not present at balance.  However, this conclusion was drawn from observations at very large imbalance, where the possibility of competing independent layer phases is large. Indeed, Clarke, {\it et al.}\cite{clarke} suggested that the system first exhibited the $(\nu_1,\nu_2)$ = (1/3, 2/3) fractional QHE before condensing into the interlayer phase coherent $\nu_T = 1$ QHE at still larger imbalances.

All of these prior studies rely upon conventional transport, i.e. with parallel currents in the two layers.  Consequently, they reflect the existence of a QHE as a transport phenomenon, but are not directly sensitive to the presence, or lack, of spontaneous interlayer phase coherence in the ground state. In addition, these earlier studies do not establish the shape of the phase boundary, in the $d/\ell$ vs. $\Delta N/N_T$ plane, separating the excitonic phase from the weakly-coupled compressible phase.  In contrast, the present measurements employ observables (tunneling and Hall drag) which involve the antisymmetric transport channel and are thus directly dependent upon interlayer phase coherence and hence allow, as Fig. 3 demonstrates, determination of the phase boundary.

The effect of layer density imbalance on the bilayer $\nu_T = 1$ quantized Hall state has been examined theoretically by several groups\cite{brey,hanna,yogesh}. Joglekar and MacDonald\cite{yogesh} offer a quantitative prediction the shape of the phase boundary.  In their Hartree-Fock theory the magneto-roton minimum in the collective mode spectrum of the strongly-coupled phase deepens as $d/\ell$ increases, signaling incipient instability against charge density wave formation.  The critical $d/\ell$ is assumed to correspond the vanishing of the magneto-roton gap. Joglekar and MacDonald find that the collective mode spectrum stiffens and the critical $d/\ell$ increases quadratically with $\Delta_v$, the splitting between the single-particle ground states in the two quantum wells. Since the interlayer capacitance in the $\nu_T = 1$ coherent phase is only slightly renormalized\cite{yogesh}, $\Delta_v$ is essentially proportional to $\Delta N/N_T$.  The dashed line in Fig. 3 shows their estimate of the phase boundary, shifted vertically by $\Delta(d/\ell) = 0.58$.  The qualitative agreement is seen to be good.

In conclusion, we have used interlayer tunneling and Coulomb drag to establish the layer density difference dependence of the phase boundary separating the interlayer coherent excitonic phase from the weakly-coupled compressible phase at $\nu_T = 1$. We find that layer density imbalance enhances the stability of the coherent phase and that the critical layer separation increases quadratically with $\Delta N/N_T$.

We thank Allan MacDonald, Yogesh Joglekar, and Kun Yang for enlightening conversations. This work was supported by the NSF under Grant No. DMR-0242946 and the DOE under Grant No. DE-FG03-99ER45766.


\begin{thebibliography}{99}
\bibitem{noise} The conductance noise floor in our experiments is about $4\times10^{-10}\Omega^{-1}\rm Hz^{-1/2}$.
\bibitem{shift} In tunneling we found that a small density imbalance, $\Delta N/N_T \approx 0.02$, developed upon application of the magnetic field required to produce $\nu_T  = 1$. As a result, all $\Delta N/N_T$ values quoted here for tunneling include a shift of +0.02.
\bibitem{tunneldrag} Tunneling between the layers can distort drag measurements, but we have demonstrated that this is not the case for the data presented here, in spite of the large enhancement of the zero bias tunneling conductance in the excitonic phase. See Ref.\cite{kellogg} for details.
\end{thebibliography}
\end{document}